The theoretical link between voltage loss, reduction in field enhancement factor, and Fowler-Nordheim-plot saturation

Richard G. Forbes[a]

Advanced Technology Institute & Department of Electrical and Electronic Engineering, University of Surrey, Guildford, Surrey, GU2 7XH, United Kingdom

With a large-area field electron emitter, when an individual post-like emitter is sufficiently resistive, and current through it sufficiently large, then voltage loss occurs along it. This Letter provides a simple analytical and conceptual demonstration that this voltage loss is directly and inextricably linked to a reduction in the field enhancement factor (FEF) at the post apex. A formula relating apex-FEF reduction to this voltage loss was obtained in the paper by E. Minoux, O. Groening, K. B. K. Teo, S. H. Dalal, L. Gangloff, J.-P. Schnell, L. Hudanski, I. Y. Y. Bu., P. Vincent, P. Legagneux, G. A. J. Amaratunga, and W. I. Milne [Nano Lett. **5**, 2135 (2005)] by fitting to numerical results from a Laplace solver. This Letter derives the same formula analytically, by using a "floating sphere" model. The analytical proof brings out the underlying physics more clearly, and shows that the effect is a general phenomenon, related to reduction in the magnitude of the surface charge in the most protruding parts of an emitter. Voltage-dependent FEF-reduction is one cause of "saturation" in Fowler-Nordheim (FN) plots. Another is a voltage-divider effect, due to measurement-circuit resistance. An integrated theory of both effects is presented. Both together, or either by itself, can cause saturation. Experimentally, if saturation occurs but voltage loss is small (< 20 V, say), then saturation is more probably due to FEF-reduction than voltage division. In this case, existing treatments of electrostatic interaction ("shielding") between closely spaced emitters may need modification. Other putative causes of saturation exist, so the present theory is a partial story. Its extension seems possible, and could lead to a more general physical understanding of the causes of FN-plot saturation.

a)Electronic mail: r.forbes@trinity.cantab.net

The last twenty years have seen much interest in possible applications of large-area field electron emitters (LAFEs), especially those based on carbon nanotubes (CNTs), or, more recently, carbon nanofibers (CNFs) (e.g., Refs 1-3]). One form of ideal LAFE can be visualised as a regular array of near-identical post-like emitters, standing upright on a flat cathode plate, called here the "emitter plate".

To model a single emitter, the "hemisphere-on-cylindrical post" (HCP) physical model is often used. This takes the emitter as a cylindrical classical conductor, of radius $r$ and length $\ell$, capped by a conducting hemisphere also of radius $r$, as illustrated in Fig. 1a. Points "a" and "c" label the emitter apex and a point on the circle of join between cap and cylinder, respectively. For simplicity, all model surfaces (both post and emitter plate) are given the same work function $\phi$.

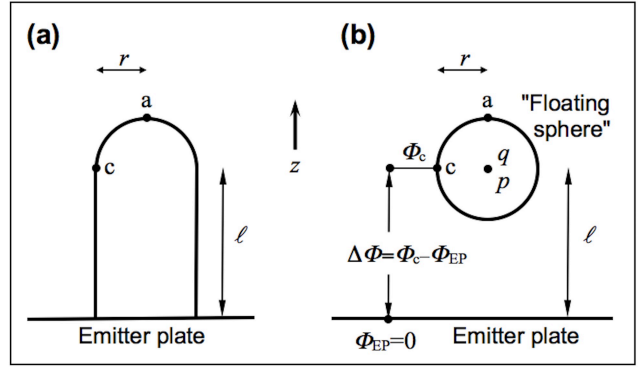

FIG. 1. (a) The hemisphere-on-cylindrical-post (HCP) physical model for a field emitting carbon nanotube or nanofiber. (b) The related "floating sphere" model used as an approximation to the HCP model. These diagrams are not to scale: normally the ratio of height $\ell$ to radius $r$ is much greater than is shown. The quantity $\Delta\Phi$ is the difference in electrostatic potential between point "c" and the emitter plate (EP). For other nomenclature, see text.

It is assumed that, in the absence of the emitter, there would be a uniform classical electrostatic field $E_M$ (the *macroscopic field*) in space above the emitter plate. In this paper, as in Ref. 4, classical electrostatics is used, and the positive field direction is taken as from the emitter plate into vacuum; thus, for a field electron emitter, values of fields and charges below are negative. As in Ref. 4, the reference zero for electrostatic potential $\Phi$ is taken to be the potential $\Phi_{EP}$ at a point on the emitter plate far distant from the emitter location.

When the HCP-model emitter is present, then the total field $E_a$ at its apex is enhanced relative to the macroscopic field, and a *macroscopic apex field enhancement factor* (FEF) $\gamma_a$ is defined by:

$$\gamma_a \equiv E_a / E_M . \qquad (1)$$

There is no known exact analytical method of determining $\gamma_a$, but many approximate analytical methods, and also numerical methods, have been used to estimate $\gamma_a$ (for example, see Refs 4-8).

Physically, nearly all these derivations take the whole emitter plate to be at constant electrostatic potential $\Phi_{EP} = 0$, and take the whole post surface to be at the same constant potential $\Phi_{EP}$. In particular, they make $\Phi_a = \Phi_c = 0$. This is equivalent to assuming that the effects of any current through the emitter may be disregarded, and that the Fermi level (and hence the thermodynamic voltage[9]) may be taken constant throughout the emitter and emitter plate. This can be called the *small-current electrostatic approximation (SCEA)*. The related apex-FEF value is denoted here by $\gamma_a^{sc}$.

However, when the emitter is sufficiently resistive and the current sufficiently high, then voltage loss ("Fermi-level variation") occurs along the post. (The term "voltage loss" refers to the fact that the voltage between the emitter apex and the counter-electrode will be less than the measured voltage provided by the high-voltage generator.) A simple analytical model is used below to show clearly that this voltage loss ($V_d$) along the post is directly and inextricably associated with a reduction in apex FEF, with both effects being caused by a reduction in the magnitude of the electric charge in the emitter apex region. This voltage loss is associated with a non-zero difference $\Delta\Phi$ in electrostatic potential between point "c" and the emitter plate.

The link between voltage loss and FEF reduction was pointed out by Groening et al.[10] in their 1999 paper. In a 2005 paper[11], Minoux et al. found a relationship between apex-FEF reduction and voltage loss along the emitter by solving Laplace's equation numerically, and fitting a formula to the results. For the case of no contact resistance at the plate/emitter interface, Minoux et al. inferred a relation that (in the notation used here) can be written

$$\gamma_a = \gamma_a^{sc} [1 - V_d/|E_M|\ell] . \qquad (2)$$

The discussion here reaches the same formula, but the simple analytical model used here brings out the underlying physics more clearly. Further, the argument here can be generalised qualitatively, to apply to emitters of any shape across which current-related voltage loss occurs.

In the SCEA, a simple method of deriving a formula for the apex FEF $\gamma_a^{sc}$ uses the so-called "Floating Sphere at Emitter Plate Potential" model, as illustrated in Fig. 1b, taking $\Delta\Phi=0$. This model, and its relation to electron thermodynamics, were recently reviewed[4]. Several levels of mathematical approximations are possible, all of which provide qualitative understanding and get qualitative trends correct, but none of which yields precisely accurate estimates of $\gamma_a^{sc}$. A particularly simple mathematical approximation (Approach II in Ref. 4) is used here to demonstrate the physics of what is happening.

The methodology behind this kind of modelling is as follows. First place charges and dipoles at appropriate locations and choose their values so that the potentials at one or two specified locations (here "a" and "c") have the values desired physically. Then use these charges (and the macroscopic field) to estimate the total field at the emitter apex "a", and hence the apex FEF.

With the SCEA, the method proceeds as follows[4-6]. First, a dipole is placed at the centre of the floating sphere, of strength $p$ such that $p/(4\pi\varepsilon_0 r^2) = rE_M$. This ensures that, when the sphere is immersed in the field $E_M$, its surface is an equipotential, and in particular that $\Phi_c = \Phi_a$. A charge is then placed at the sphere centre, of strength $q^{sc}$ such that

$$\Delta \Phi \equiv (\Phi_c - \Phi_{EP}) = q^{sc}/4\pi\varepsilon_0 r - E_M \ell = 0 . \tag{3}$$

This ensures that the floating sphere is at potential $\Phi_{EP}$ (if image contributions are neglected). At the emitter apex, this "sphere charge" $q^{sc}$ creates a field contribution $E_{a,q}$ given, using (3), by

$$E_{a,q} = q^{sc}/4\pi\varepsilon_0 r^2 = (\ell/r) E_M . \tag{4}$$

In a fuller treatment[4,6] there would be other contributions, associated with the images of the sphere charge and dipole in the emitter plate, but—for typical experimental values of the ratio $(\ell/r)$—often 100 or more—the resulting contributions to the total apex field $E_a$ are negligible in comparison with the sphere contribution $E_{a,q}$. This leaves just the contributions to $E_a$ resulting from the sphere charge, the sphere dipole and the macroscopic field. However, under normal circumstances, the $E_{a,q}$ term is much larger than the other two (see Ref. 4) and we reach the well-known result

$$\gamma_a^{sc} = E_a/E_M \approx E_{a,q}/E_M = (\ell/r) . \tag{5}$$

In the case where significant current flows, and conditions are such that a significant voltage loss $V_d$ occurs, the above treatment is easily modified, as follows. For simplicity, it is assumed that the whole voltage loss occurs along the cylinder of the HCP model. In classical electromagnetism, the apex of a field electron emitter is more positive that the emitter plate, and $V_d$ is positive. Because the work function is taken the same for all surfaces, this corresponds to an electrostatic potential difference (PD) $\Delta\Phi$ along the cylinder surface given by $\Delta\Phi = V_d$. It is useful to write $\Delta\Phi$ as a fraction $k$ of the (positive) PD $(-E_M\ell)$ induced between "c" and the emitter plate by the macroscopic field; hence

$$\Delta\Phi = V_d = -kE_M\ell . \tag{6}$$

To allow this PD to be present in the floating-sphere model, the magnitude of the (negative) sphere charge has to be reduced. Equation (3) above has to be replaced, and the sphere charge needs to be changed (by a positive amount $\Delta q$) from $q^{sc}$ to the value $q$ given via

$$\Delta \Phi = q/4\pi\varepsilon_0 r - E_M \ell = -kE_M \ell, \qquad (7)$$

$$q/4\pi\varepsilon_0 r = (1-k)E_M \ell . \qquad (8)$$

From the same argument as before, that the apex field $E_a$ is dominated by the field contribution due to the sphere charge $q$ (provided, in this case, that $k$ is not very close to unity) it follows that, when the SCEA is abandoned, the apex field and FEF are normally given adequately by

$$E_a \approx q/4\pi\varepsilon_0 r^2 = (1-k)(\ell/r) E_M \approx (1-k) \gamma_a^{sc} E_M , \qquad (9)$$

$$\gamma_a = E_a / E_M \approx (1-k) \gamma_a^{sc} = \gamma_a^{sc} \{1 - V_d/(-E_M \ell)\} \equiv \Theta_{fr} \gamma_a^{sc} . \qquad (10)$$

where the correction factor $\Theta_{fr}$ [$= \gamma_a/\gamma_a^{sc} = 1 - V_d/(-E_M \ell)$] is to be attributed to "FEF reduction (fr)".

Since $E_M$ is negative for a field electron emitter, eq. (10) is the same result as eq. (2) above found[11] by fitting to numerical simulation. However, it is explicitly clear here that the sphere-charge $q$ appears in expressions both for the apex field and for the potential difference between the cylinder ends, and hence for the "voltage loss along the emitter". It follows that FEF reduction and current-induced voltage loss along the emitter are directly and inextricably linked, physically.

The use of a simple analytical model has led to an explicit formula. However, it is clear that—qualitatively—the effect is a general one, applicable to an emitting protrusion of any shape, and not dependent on the size of the emitter apex or on the precise local geometry or nature of emission sites. When current through the protrusion leads to significant voltage loss (in accordance with Ohm's law), then this voltage loss is associated with a reduction in the magnitude of the charge in the most vacuum facing parts of the protrusion, and hence with reductions in the magnitude of the local barrier field and FEF there.

Although the floating-sphere model allows the basic physics of current-induced FEF-reduction to be displayed, allows eq. (2) to be retrieved, and allows the above qualitative conclusions to be drawn, it needs to be emphasised that it is not (and is not intended to be) a quantitatively accurate model. In particular, it will not deal accurately with situations where the emitter is cone-shaped rather than post-shaped, or where the electrical resistance is non-uniformly distributed (as would occur if most of the resistance is across a poor contact between the emitter and the substrate). In such cases, more sophisticated modelling in needed. Thus, Minoux et al. find[11] that, when most of the resistance is in the contact, a correction factor $\alpha$ has to be included, and the r.h.s. of eq. (2) becomes $\gamma_a^{sc}[1-\alpha V_d/|E_M|\ell]$ (they find $\alpha=0.92$).

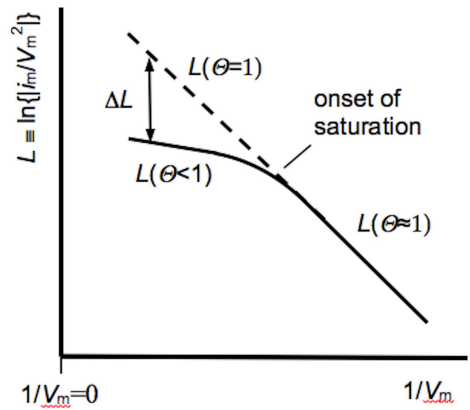

FIG. 2. Schematic diagram that illustrates the definition of the (positive) quantity $\Delta L$ given by eq. (15).

A practical context in which FEF-reduction issues arise is the explanation of "saturation" in Fowler-Nordheim (FN) plots, which causes the plots to adopt the kinked form illustrated schematically in Fig. 2. It is readily shown that eq. (10) leads to this effect. This is true for any Fowler-Nordheim-type (FN-type) equation. The exponent $-G^{GB}$ of the "general form" FN-type equation[12] is

$$-G^{\mathrm{GB}} = -\nu_{\mathrm{F}}^{\mathrm{GB}} b\phi^{3/2}/|E_{\mathrm{a}}|, \tag{11}$$

where $b$ is the second FN constant[13], and $\nu_{\mathrm{F}}^{\mathrm{GB}}$ is the relevant barrier form correction factor[12].

The (negative) macroscopic field $E_{\mathrm{M}}$ is related to the (positive) voltage $V_{\mathrm{p}}$ applied between a counter-electrode ("anode", with field electron emission) and the emitter plate, by $E_{\mathrm{M}} = -V_{\mathrm{p}}/\zeta_{\mathrm{M}}$. Here, $\zeta_{\mathrm{M}}$ is the relevant (positive) macroscopic conversion length. In a given practical arrangement, $\zeta_{\mathrm{M}}$ is a constant that depends on the system geometry; in planar-parallel-plate geometry $\zeta_{\mathrm{M}}$ is adequately given by the plate separation. Thus, the relationship between $E_{\mathrm{a}}$ and $V_{\mathrm{p}}$ becomes

$$|E_{\mathrm{a}}| = (\gamma_{\mathrm{a}}/\zeta_{\mathrm{M}})V_{\mathrm{p}} = (\Theta_{\mathrm{fr}}\gamma_{\mathrm{a}}^{\mathrm{sc}}/\zeta_{\mathrm{M}})V_{\mathrm{p}}. \tag{12}$$

A complication arises if the emitter plate is itself sufficiently resistive that the voltage $V_{\mathrm{p}}$ between the plate front surface (facing vacuum) and the anode is not equal to the measured voltage $V_{\mathrm{m}}$. In this case, a "voltage divider (vd) effect" will occur[12], and the two voltages will be related by $V_{\mathrm{p}} = \Theta_{\mathrm{vd}}V_{\mathrm{m}}$, where $\Theta_{\mathrm{vd}}$ is a correction factor that is current dependent and will lie in the range $0<\Theta_{\mathrm{vd}}\leq 1$. In this more general case, the relationship between $E_{\mathrm{a}}$ and $V_{\mathrm{m}}$ can be written

$$|E_{\mathrm{a}}| = (\Theta\gamma_{\mathrm{a}}^{\mathrm{sc}}/\zeta_{\mathrm{M}})V_{\mathrm{m}} \equiv c_{\mathrm{m}}V_{\mathrm{m}}, \tag{13}$$

where the total correction factor $\Theta = \Theta_{\mathrm{vd}}\Theta_{\mathrm{fr}}$, and $c_{\mathrm{m}}$ [$\equiv \Theta\gamma_{\mathrm{a}}^{\mathrm{sc}}/\zeta_{\mathrm{M}}$] is an auxiliary parameter introduced to simplify equation presentation. It follows that, in FN coordinates of type $\ln\{|i_{\mathrm{m}}|/V_{\mathrm{m}}^2\}$ vs $1/V_{\mathrm{m}}$, the general-form FN-type equation[12] can be written

$$L \equiv \ln\{|i_{\mathrm{m}}|/V_{\mathrm{m}}^2\} = \ln\{A_{\mathrm{f}} a\phi^{-1}c_{\mathrm{m}}^2\} - \nu_{\mathrm{F}}^{\mathrm{GB}}b\phi^{3/2}/c_{\mathrm{m}}V_{\mathrm{m}}, \tag{14}$$

where $a$ is the first FN constant[13], and $A_{\mathrm{f}}$ is the emitter's formal emission area[12] as defined in the context of this equation.

In circumstances where $\Theta$ effectively has the value 1 (which is usually the case for low measured voltages), $c_m$ has the constant value $c_m^{sc} = \gamma_a^{sc}/\zeta_M$. However, from eq. (12), if at higher measured voltages, $\Theta$ becomes progressively smaller, then $c_m$ will become progressively smaller. As illustrated in Fig. 2, let $\Delta L$ denote the difference between the quantity $L(\Theta=1)$ evaluated using the SCEA and the quantity $L(\Theta<1)$. It is readily shown that $\Delta L$ is predicted adequately by:

$$\Delta L \approx 2\ln\{c_m^{sc}/c_m\} + (v_F^{GB} b\phi^{3/2}/c_m^{sc} V_m)\{(c_m^{sc}/c_m)-1\} . \qquad (15)$$

Clearly, as $V_m$ increases (and $1/V_m$ decreases), then $c_m$ progressively decreases below $c_m^{sc}$, and $(c_m^{sc}/c_m)$ becomes progressively greater than 1. The effect is that both terms in eq. (15), and hence $\Delta L$, progressively increase as $(1/V_m)$ decreases. This is the effect often called "saturation". The actual calculation of values for $c_m$ (or of the shape of saturated FN plots) is non-trivial and is outside the scope of this letter.

The treatment here, in terms of $\Theta=\Theta_{vd}\Theta_{fr}$, shows that saturation can be caused either by a voltage-divider effect or by FEF reduction, or by both acting together. Both effects are due to significant series resistance in the measurement circuit. The former occurs when the resistance is in the emitter plate itself (for example, if the plate is poorly conducting silicon, or if some other form of ballast resistance has been designed in), the latter when the resistance is in the emitting protrusion or the contact between protrusion and plate.

Most past discussions of series-resistance effects on FN plots have concentrated on the voltage-divider explanation. However, simulations of voltage-divider effects by the present author and Deane (see Ref. 12) did not lead to plausible-looking simulated plots. By contrast, the simulations[11] of Minoux et al, using the FEF-reduction explanation, did generate plausible-looking FN plots. This leads the author to suspect that, for LAFEs, FEF reduction may usually be the more plausible explanation.

The following illustrative argument is a further indicator. Consider a resistive emitting post with $\phi$=4.5 eV, radius $r$= 10 nm, length $\ell$ = 1 μm, and take the small-current FEF $\gamma_a^{sc}$ to be 100. Such a post would field emit significantly at a barrier field $E_a$ of around –4 V/nm, and (if no series-resistance effects occur) a macroscopic field $E_M$ around –40 V/μm. The related value of $(-E_M \ell)$ would be 40 V. To get a value $\Theta_{fr}$=0.5, one would need a voltage loss (along the post) of 20 V. By contrast, an illustrative value for $V_m$ (assuming a typical value of $\zeta_M$ as 25 μm) might be 2000 V. To get a value $\Theta_{vd}$=0.5 by the voltage-divider effect, one would need a voltage loss of 1500 V across the series resistance associated with the emitter plate and the rest of the path to the high-voltage generator.

The large difference between 1500 V and 20 V makes it look plausible that, in many cases where series resistance can be presumed responsible for saturation, and especially with LAFEs, the detailed cause is more likely to be FEF reduction than the voltage-divider effect.

On the other hand, with other situations, for example flash memory devices where electrons are field-injected into an oxide layer, presumably from metallic nanoprotrusions on metal electrodes, the voltage-divider explanation may look more relevant.[14]

Of course, field electron emission into vacuum from metallic emitters with a good conducting path to the high-voltage generator represents a situation where effectively $\Theta$=1 over the whole working range, and the SCEA applies throughout the range.

Experimental estimates of the total voltage loss across the whole series resistance are possible in principle, either by putting a metal probe in direct contact with the emitter, or by retarding potential energy analysis of field emitted electrons (e.g., Refs. 10,15,16). The measurement of a relatively small total voltage loss associated with the saturated part of the $i_m(V_m)$ characteristics would be an indication that saturation is probably due to a FEF-reduction effect. Minoux et al. used a direct-contact probe in an experiment of this type, finding $V_d$ values between 0 and 4 V.

With LAFEs, the quantitative theory above applies to single emitters that do not significantly interact electrostatically with adjacent emitters, because they are sufficiently well spaced. However, FEF-reduction effects will also occur when electrostatic interaction (usually called "shielding") takes place between emitters. Recently, many papers discussed electrostatic interactions between closely-spaced emitters (e.g., Refs 4,7,8), all using the SCEA. In physical situations where the SCEA is not valid, numerical results relating to electrostatic interactions may need modifying.

The theory above does not cover all putative causes of saturation-like effects. Other putative causes include voltage-dependent relative changes in work-function or operative work function, either as a result of adsorbate behaviour (perhaps influenced by joule heating), or—with semiconductors—as a result of the kind of field penetration and band-bending effects that occur in the Modinos[17] "zero-current approximation". Hopefully it may be possible, as some future point, to extend the present theory to cover some or all of these effects, by relaxing the constant-work-function assumption made (for simplicity) in this Letter.

Finally, I suggest that kinked FN plots as indicated in Fig, 2, should be regarded, not as evidence of some sort of anomaly, but as the actual current-voltage characteristics (presented in FN coordinates) of an electronic circuit device that is basically a form of diode. If you need a device with different characteristics, then try modifying its structure or the materials from which it is built. (In fact, this is what was done by Minoux et al., who eliminated their voltage-loss and FEF-reduction effects by thermal processing of their carbon nanotubes, thereby reducing their resistivity.)

I thank the University of Surrey for provision of facilities, and thank Professor S.R.P. Silva for reinforcing my view that field electron emitters, especially LAFEs, could usefully be thought of as electronic-circuit elements with measured current-voltage characteristics.